# Geospatial Analysis Requires a Different Way of Thinking: The Problem of Spatial Heterogeneity

Bin Jiang

Department of Technology and Built Environment, Division of Geomatics
University of Gävle, SE-801 76 Gävle, Sweden
Email: bin.jiang@hig.se

*(Draft: May 2013, Revision: July 2013, Janurary, February 2014)*

**Abstract**
Geospatial analysis is very much dominated by a Gaussian way of thinking, which assumes that things in the world can be characterized by a well-defined mean, i.e., things are more or less similar in size. However, this assumption is not always valid. In fact, many things in the world lack a well-defined mean, and therefore there are far more small things than large ones. This paper attempts to argue that geospatial analysis requires a different way of thinking - a Paretian way of thinking that underlies skewed distribution such as power laws, Pareto and lognormal distributions. I review two properties of spatial dependence and spatial heterogeneity, and point out that the notion of spatial heterogeneity in current spatial statistics is only used to characterize local variance of spatial dependence. I subsequently argue for a broad perspective on spatial heterogeneity, and suggest it be formulated as a scaling law. I further discuss the implications of Paretian thinking and the scaling law for better understanding of geographic forms and processes, in particular while facing massive amounts of social media data. In the spirit of Paretian thinking, geospatial analysis should seek to simulate geographic events and phenomena from the bottom up rather than correlations as guided by Gaussian thinking.

**Keywords:** Big data, scaling of geographic space, head/tail breaks, power laws, heavy-tailed distributions

## 1. Introduction

Geospatial analysis, or spatial statistics in particular, has been dominated by a Gaussian way of thinking, which assumes that things are more or less similar in size, and can be characterized by a well-behaved mean. Based on this assumption, extremes are rare; if extremes do exist, they can be mathematically transformed into normal things (e.g., by taking logarithms or square roots). This Gaussian thinking is widespread, and has dominated the sciences for a very long time. However, Gaussian thinking has been challenged and been accused of misrepresenting our world (Mandelbrot and Hudson 2004; Taleb 2007). Indeed, many things in the world are not well behaved or lack of a well-behaved mean. This can seen from the extreme events such as the September 11 attacks. The extent of devastation of such events was enormous and beyond any predictions and estimations. This is the same for many geographic features, which exhibit a pretty skewed or heavy-tailed distribution such as power laws and lognormal distributions. The heavy-tailed distributions imply that there are far more small geographic features than large ones, namely scaling of geographic space.

A power law distribution is often referred to as scale free, literally meaning a lack of average for characterizing the sizes of things (Barabási and Albert 1999). The power law distribution has been given different formats for it was discovered by different scientists in different disciplines over the past 100 years. Among several alternatives, Zipf's law (Zipf 1949) and the Pareto distribution (Pareto 1897) are the two formats most frequently referred to in the literature. Zipf's law, with respect to city sizes, implies that there are far more small cities than large ones, while the Pareto distribution indicates that there are far more poor people than rich people, or equivalently far more ordinary people than extraordinary people. The Pareto distribution has been popularized as the 80/20 principle (Koch 1999) or the long tail theory (Anderson 2006) in the popular science and business literature. The heavy-tailed distribution, including power laws, lognormal and others similar, is what underlies the new way of thinking I want to advocate in this paper. The central argument is that geospatial analysis requires a new way of thinking radically different from Gaussian thinking, and spatial heterogeneity should be formulated as a scaling law of geography.

This is an unprecedented time when we face increasing rich geographic data sources based not only on the legacy of traditional cartography and remote sensing imagery, but also emerging from various social media such as Flickr, Twitter, and OpenStreetMap, collectively known as volunteered geographic information (Goodchild 2007). Today, one can amass gigabytes of an entire country's data for geospatial analysis and computing, for both data volumes and computing capacity have increased dramatically. However, our mindsets, subsequently our analysis methods, have been relatively slower to adapt the rapid changes (Mayer-Schonberger and Cukier 2013). For example, we tend to sample data rather than take all data for geospatial analysis; we tend to transform skewed data into "normal" by taking logarithms for example. The sampling and logarithm transformation have distorted the underlying property of the data before the data can yield insights. The old way of thinking, Gaussian thinking, that relies on a well-defined mean to characterize geographic features, is a major barrier to achieving deep and new insights into geographic forms and processes.

Many analytical techniques have been developed, in both standard and spatial statistics, to address outliers, to measure skewness and autocorrelation, and to test significance. However, what I want to argue in this paper is not these techniques per se, but something radical in the way of thinking. Gaussian thinking, based on the assumption of independent things in a simple, static, and equilibrium world, is essentially a typical linear thinking, which implies that small cause small effect, large cause large effect, and the whole is equl to the sum of its parts. This linear thinking is a simple way of thinking guided by the reductionism philosophy, and for understanding a simple world in essence (see Section 2.2 for more details). The reader may argue that spatial statistics differs from standard statistics in spatial dependence or spatial autocorrelation. It is indeed true, but the notion of spatial dependence or autocorrelation does not help us to go beyond Gaussian thinking assumed by standard statistics, for we tend to characterize things by a well-defined mean with a limited variance. It is well recognized that geographic forms are fractal rather than Euclidean, and geographic processes are nonlinear rather than linear (Batty and Longley 1994, Chen 2009). In other words, a geographic system is a complex nonlinear world, in which there is the butterfly effect, and the whole is greater than the sum of its parts. In this paper, I attempt to argue that the Paretian way of thinking, founded on the assumption of interdependent things in a complex, dynamic, and nonequilibrium world, is more appropriate for geospatial analysis, and for better understanding geographic forms and processes. Geospatial analysis, while facing increasing amounts of social media data, should seek to uncover the underlying mechanisms through simulations from the bottom up rather than simple causality or correlations.

The remainder of this paper is organized as follows. Section 2 introduces, in a pedagogic manner, two distinct statistic distributions, namely Gaussian- and Paretian-like distributions, with a particular focus on the underlying ways of thinking. Section 3 reviews two unique properties of spatial dependence and spatial heterogeneity, and points out that the notion of spatial heterogeneity in current spatial statistics is only used to characterize local variance of spatial dependence. I therefore argue, in Section 4, that spatial heterogeneity should be formulated as a scaling law, and suggest some effective ways of detecting and revealing the scaling law and pattern for geographic features. I further discuss, in Section 5, some deep implications of Paretian thinking and the scaling law before draw a summary in Section 6.

## 2. Two distint distributions and the underlying ways of thinking
In this section, I first illustrate statistical differences between a homogenous Gaussian-like distribution and a heterogeneous Paretian-like distribution (Note the 'homogenous' is relative to the 'heterogeneous'; see Section 2.1 for more details), using temperature and population of major US cities, and based respectively on histograms and rank-size plots. The temperature is the annual average maximum during 1981-2010, taken from the site: http://www.prism.oregonstate.edu/products/matrix.phtml, while the population is according to the 2010 US census. I then elaborate on the underlying ways of thinking or world views associated with the two categories of distributions.

**2.1 Gaussian- versus Paretian-like distributions**
If we carefully examine two variables – temperature and population – of 720 major U.S. cities with population greater than 50,000 people, we can see that the two variables are very distinct. Although not a normal distribution, the temperature can be well characterized by its mean 20.6 (Figure 1a). One can estimate a city's temperature fairly accurate and precise based on the mean value, since the highest is 31.6, and the lowest is 9.3. In other words, the mean 20.6 is a typical temperature for US cities. The distribution that can be characterized by a well-defined mean is referred to as a Gaussian-like distribution including for example the binomial and Poisson distributions. This temperature distribution can be further assessed from the detailed statistics as shown in Table 1 (the temperature column). The range between the highest (31.6) and the lowest (9.3) is not very big (22.3), and the ratio of the highest to the lowest is as little as 3.4. The two measures of central tendency – mean and median – are the same. The standard deviation is 4.9, about one quarter of the range. This statistical picture of the temperature is very distint from that of the city size or population.

The histogram of the population is extremely right skewed (Figure 1b). This extreme skewness is reflected in several parameters: a wide range (8,273,676), a huge ratio (166), and a large standard deviation (393,004); see the population column of Table 1. In such a significantly skewed distribution, the mean of 157,467 make little sense for characterizing the population. In other words, the mean of 157,467 does not represent a typical size of the U.S. cities, since the largest city is as big as 8 millions, while the smallest city is as small as 50 thousands. The right skewed histogram indicates that there are far more small cities than large ones in the U.S. No wonder that the two mesures of central tendency – mean and median - differ from each other significantly; refer to Table 1 (the population column) for more details. The standard statistics, or the histogram in particular, is little effective for describing data with a heavy-tailed distribution such as city sizes. Instead, power law based statistics, or rank-size plots in particular, should be adopted for characterizing this kind of data.

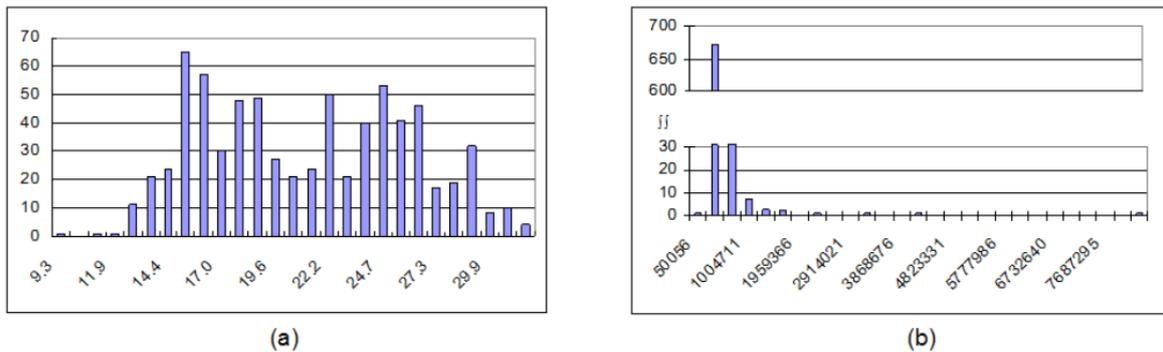

Figure 1: (Color online) Histograms of (a) the temperature, and (b) the population of U.S. cities (Note: the two distint distributions indicate respectively Gaussian-like and Paretian-like distributions.)

Table 1: Statistics about temperature and population of U.S. cities

| Statistics | Temperature | Population |
|---|---|---|
| Minimum | 9.3 | 50,056 |
| Maximum | 31.6 | 8,323,732 |
| Range | 22.3 | 8,273,676 |
| Ratio | 3.4 | 166 |
| Mean | 20.6 | 157,467 |
| Median | 20.6 | 82,115 |
| Mode | 14.9 | 62,820 |
| St. Dev. | 4.9 | 393,004 |

Instead of plotting temperature and population on the x-axis (as in the histograms), they are plotted on the y-axis, while the x-axis is the ranking order. This way of plotting is called rank-size plot, or rank-size distribution (Zipf 1949). The largest city (in terms of population) ranks number one, followed by the second largest, and so on. The same arrangement is made for the temperature; the highest temperature city ranks number one, followed by the second highest, and so on. The two distribution curves look very different; the temperature curve drops gradually, and then reaches quickly the minimum, while the population curve drops quickly and then gradually reaches the minimum (Figure 2). Note that the red parts in the figure are those above the averages, called the head, while those below the averages, called the tail, are shown in blue. More specifically, 362 cities (approximately 50 percent) are above the average temperature 20.6, while only 146 cities (approximately 20 percent) are above the average city size 157,467. Clearly, a heavy or long tail (80 percent in the tail) exists for the population distribution, but a short tail (50 percent) for the temperature distribution. Generally, a heavy-tailed distribution possesses an inbuilt imbalance between the head and the tail (e.g., a 70/30 or 80/20 relationship). This imbalance indicates a nonlinear relationship between the head and the tail. Such an inbuilt imbalance, or nonlinearity, is clearly missing in a Gaussian-like distribution with a well-balanced relationship between the head and the tail (e.g., 50/50).

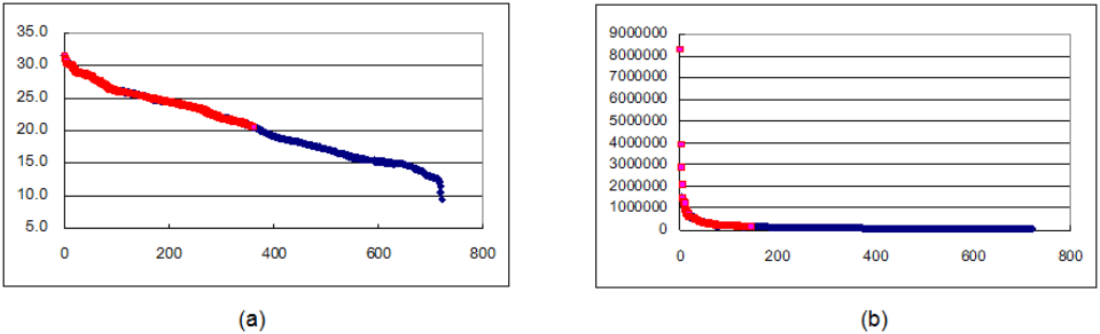

Figure 2: (Color online) Rank-size plots of (a) the temperature, and (b) the population of the U.S. cities
(Note: Values above and below the averages are respectively in red and blue; clearly, there is a short head and a long tail for the population, forming an unbalanced contrast, while the values above and below the averages are more or less the same for the temperature.)

**2.2 The underlying ways of thinking**
The differences between the two distributions lie fundamentally in different ways of thinking, or different ways of viewing the world, rather than different techniques associated with each distribution. Technically, data with a Paretian-like distribution can be easily transformed into a Gaussian-like distribution, e.g., by taking logarithms. Gaussian thinking implies more or less similar things in a simple, static, and equilibrium world, while Paretian thinking believes in far more small things than large ones in a dynamic, complex, and nonequilibrium world (McKelvey and Andriani 2005; see Table 2). Standard statistics teaches us that if the probability of an event is small, then the event occurs rarely. The event can be considered an outlier that is literally distant from the rest of the data. However, in Paretian thinking, an event of small probability, or the highly improbable, has a significant impact (e.g., the September 11 attacks) and thus be ranked highly.

Table 2: Comparison between the two ways of thinking

| Gaussian thinking | Paretian thinking |
|---|---|
| With a mean (or scale) | Without a mean (or scale-free) |
| Static | Dynamic |
| Simple | Complex |
| Equilibrium | Non-equilibrium |
| Linear | Nonlinear |
| Predictable | Unpredictable |

In Gaussian thinking, the world does not change much, and all changes occur around a stable and well-defined mean. Thus, the presumed Gaussian world is static, simple, linear, and predictable. The Newtonian physics is sufficient to understand and deal with the Gaussian world. Why does such a predictable world exist? Such a world reflects a lack of interaction and competition among individual agents; every agent acts independently without influences upon or affects from others. This assumption is fundamental to standard statistics, and of course appropriate for many events in the world like human heights. Spatial statisics differentiate it from standard statistics in spatial dependence, but it does not change fundamentally the underlying way of thinking - Gaussain thinking with a well-defined mean for characterizing things. On the other hand, in Paretian thinking, the world is full of surprises, and changes are often dramatic and unexpected. Thus, there is no stable and well-defined mean for characterizing the surprises and changes. The presumed Paretian world is essentially dynamic, complex, nonlinear, and unpredictable. This unpredictable world is founded on the assumption that everything is related to, or interdependent with, everything else. This interdependence assumption implies that cooperation and competition would eventually lead to unbalanced results characterized by a long-tail distribution (c.f. Section 3 for more discussions).

Nature is awash with phenomena such as trees, rivers, mountains, clouds, coastlines, and earthquakes that exhibit power laws or heavy-tailed distributions in general (Mandelbrot 1982, Schroeder 1991, Bak 1996). Accordingly, power law has been formulated as a fundamental law in various disciplines such as physics, biology, economics, computer science, and linguistics. People's daily activities are also governed by power laws (Barabási 2010), indicating bursty behaviors of human mobility or activities in general. Power laws are a signature of complex systems that are evolved in nonlinear manners, i.e., small causes often have disproptional large effects. For instance, the top 10 percent of the most connected streets account for 90 percent of traffic flows (Jiang 2009). In a 21-block area of Philadelphia, 70 percent of the marriages occurred between people who lived no more than 30 percent of that distance apart (Zipf 1949).

The examination of the two ways of thinking suggests that Paretian-like distribution, or Paretian thinking in general, appears more appropriate for understanding geographic forms and processes, for dependence is a key property of spatial statistics (c.f., Section 3 for more details). In spite of spatial dependence being its key property, spatial statistics is still unfortunately very much dominated by Gaussian thinking. The very notion of spatial heterogeneity refers to local variance of spatial dependence, but from global to local, or from one single correlation coefficient to multiple coefficients (c.f., Section 3 for more details). In the remainder of this paper, I review two spatial properties of dependence and heterogeneity, and argue that spatial heterogeneity is ubiquitous, and it should be formulated as a scaling law. And I further discuss some deep implications of the scaling law and Paretian thinking for better understanding of geographic forms and processes in the era of big data.

**3. Spatial properties of dependence and heterogeneity**
It is well known that in contrast to the independence assumption of standard statistics, geographic phenomena or events are not random or independent. Geographic events are more likely to occur in some locations than others (spatial heterogeneity), and nearby events are more similar than distant events (spatial dependence). Both spatial heterogeneity and spatial dependence are referred to as spatial properties, indicating respectively that geographic events are related to their locations and to their neighboring events. Spatial dependence is widely known or formulated as the first law of geography: *"Everything is related to everything else, but near things are more related than distant things"* (Tobler 1970). For example, your housing price is likely to be similar (positive correlation) to those of your neighbors. Similarly, the elevations of two locations 10 meters apart are likely to be more similar than the elevations of two locations of 100 meters apart. Note that "likely" indicates a statistical rather than a deterministic property; one can always find exceptions in statistical trends.

Spatial heterogeneity refers to no average location that can characterize the Earth's surface (Anselin 1989, Goodchild 2004). This is indeed true, while for example referring to the diversity of landscapes and species (animals and plants) on the Earth's surface (Wu and Li 2006, Bonner 2006). This diversity or heterogeneity indicates uneven geographic and statistical distributions involving both landscapes and species – that is, a mix of concentrations of multiple species (biological), terrain formations (geological), environmental characteristics (such as rainfall, temperature, and wind) on the one hand, and various concentrations of various types of species on the other. A variety of habitats such as different topographies, soil types, and climates can accommodate a greater number of species. These are the natural environments of the Earth's surface. Spatial heterogeneity in geography also concerns human-made built environments created by human activities such as industrialization and urbanization, and in particular, for example, the diversity of human settlements or cities in particular. Given the diversity or spatial heterogeneity of the Earth's surface, homogeneous Gaussian-like distribution is unlikely to be the right means to characterize complex geographic features.

Spatial dependence and spatial heterogeneity are unique properties to spatial data and geospatial analysis (Anselin 1989, Griffith 2003), and probably the two most important principles of geographic information science (GIScience). Goodchild (2004) has been a key advocator for formulating general principles for GIScience. On several occasions, he has made insightful remarks on spatial heterogeneity or spatial properties in general. His definition of spatial heterogeneity as "no average location" is in effect the notion of scale-free used to characterize things that exhibit a power law or heavy-tailed distribution (Barabási and Albert 1999). On the other hand, he stated, with respect to spatial heterogeneity, that all locations are unique, due to which geography might be better considered as an idiographic science, studying the unique properties of places (Goodchild 2004). However, I argue, in contrast to Goodchild, that spatial heterogeneity makes geography a nomothetic science. This is because spatial heterogeneity itself is a law - the scaling law, implying that there are far more small geographic features than large ones. Spatial heterogeneity is a kind of hidden order, which appears disordered on the surface, but possesses a deep order beneath. This kind of hidden order can be characterized by a power law or a heavy-tailed distribution in general.

Current spatial statistics suffers from what I call 'spatial heterogeneity paradox'. Spatial heterogeneity is defined as no average location, but we tend to use a well-defined mean or average to characterize locations. This paradox implies that our mindsets are still constrained by Gaussian thinking. The current notion of spatial heterogeneity refers to local variance of spatial dependence. This can be seen from the development of local spatial statistics and local statistical models that initially brought spatial heterogeneity into spatial statistics (Anselin 1989). Local spatial statistics concern local variants of spatial autocorrelation, a measure to spatial dependence, including, for example, the local statistical models (Getis and Ord 1992), the LISA techniques (Anselin 1995), and geographically-weighted regression (Fotheringham et al. 2002). The shifting perspective of spatial autocorrelation from global to local brings new insights into spatial dependence, or the heterogeneity of spatial dependence. However, all these techniques and models are essentially based on Gaussian statistics, using a well-defined mean with a limited variance. To paraphrase Mandelbrot (Mandelbrot and Hudson 2004), spatial heterogeneity refers to 'wild' variances, but Gaussian-like distribution can only characterize 'mild' variances.

Human activities are the major forces behind spatial heterogeneity in the built environments. While carrying out activities, human beings (and their interventions) must respect the spatial heterogeneity of nature – that is, harmonize with rather than damage the natural environments. Geographic information concerning urban and human geography captures essentially spatial variations of the built environments, which demonstrate 'wild' heterogeneity as well. For example, Zipf's law on city sizes (Zipf 1949) mainly concerns such a spatial variation. Thus, I argue, in contrast to the conventional view, that dependence, or more precisely interdependence, is a first-order effect, while heterogeneity is a second-order effect. Let us do a thought experiment. Imagine that once upon a time, there were no cities, only scattered villages. Over time, large cities gradually emerge through the interactions of villages, so do mega cities through the interactions of cities. The interactions (competition and cooperation) of villages and cities are actually those of people acting individually and/or collectively. These interactions are what I mean by dependence and interdependence. Eventually, there are far more small cities than large ones through for example the mechanism of *"the rich get richer."* This observation is the same for the wealth distribution among individuals in a country; far more poor people than rich people, or far more ordinary people tha extraordinary people (Epstein and Axtell 1996). The interactions among people and cities reflect the interdependence effect in the formation and evolution of cities and city systems, and the built environments in general.

**4. Spatial heterogeneity as a scaling law**
The subtitle of this paper *'The Problem of Spatial Heterogeneity'* is an homage to the classic work *'The Problem of Spatial Autocorrelation' (*Cliff and Ord 1969), which popularized the concept of spatial dependence. Similarly, spatial heterogeneity under Gaussian thinking is indeed a problem because it cannot be characterized by a well-defined mean. However, in the Paretian way of thinking, spatial heterogeneity is not a problem, but the norm. Spatial heterogeneity should be formulated as a scaling law in geography.

**4.1 Ubiquity of the scaling law in geography**
Geographic features are unevenly or abnormally distributed, so the scaling pattern of far more small things than large ones is widespread in geography (Pumain 2006). The scaling pattern has another name called fractal (Mandelbrot 1983). Fractal-related research in geography has concentrated too much on concepts such as fractal dimension and self-similarity. In fact, the scaling law is fundamental to all of these concepts. In this regard, Salingaros and West (1999) formulated a universal rule for city artifacts; there are far more small city artifacts than large ones, due to which the image of the city can be formed in human minds (Jiang 2013b). With the increasing availability of geographic information, the scaling law has been observed and examined in a wide range of geographic phenomena including, for example, street lengths and connectivity (Carvalho and Penn 2004, Jiang 2009), building heights (Batty et al. 2008), street blocks (Lämmer et al. 2006, Jiang and Liu 2012), population densities (Schaefer and Mahoney 2003, Kyriakidou et al. 2011), and airport sizes and connectivity (Guimerà et al. 2005). Interestingly, the scaling of geographic space has had an enormous effect on human activities; human activities and interactions in geographic space exhibit power law distributions as well (Brockmann et al. 2006, Gonzalez et al. 2008, Jiang et al. 2009). Table 3 provides a synoptic view of the ubiquity of power laws in geography, noting that the references listed are non-exhaustive, but for example only.

Table 3: Power laws in geographic features or phenomena

| Geographic phenomena | References (for example) |
| --- | --- |
| City sizes | Zipf 1949, Krugman 1996, Jiang and Jia 2011 |
| Fractals in cities or geographic space | Goodchild and Mark 1987, Batty and Longley 1994 |
| Coast lines and mountains | Mandelbrot 1967, Bak 1999 |
| Hydrological networks | Hack 1957, Horton 1945, Maritan et al. 1996, Pelletier 1999 |
| Urban and architectural space | Salingaros and West 1999 |
| Street lengths and connectivity | Carvalho and Penn 2004, Jiang 2009 |
| Building heights | Batty et al. 2008 |
| Street blocks | Lämmer 2006, Jiang and Liu 2012 |
| Population density | Schaefer and Mahoney 2003, Kyriakidou et al. 2011 |
| Airport sizes and connectivity | Guimerà et al. 2005 |
| Human mobility | Brockmann et al. 2006, Gonzalez et al. 2008, Jiang et al. 2009 |

Despite its ubiquity, ironically the scaling law, or the Paretian way of thinking in general, has not been well received in geospatial analysis as elaborated earlier in the text. Current geospatial analysis adopts a well-defined mean or average to characterize spatial heterogeneity. The two closely related concepts of scale and scaling must be comprehended together, i.e., many different scales, ranging from the smallest to the largest, form a scaling hierarchy. This comprehension should be added, as a fourth one, into the three meanings of scale in geography: cartographic, analysis, and phenomenon (Montello 2001). The essence of power laws is the scaling pattern, in which there are far more small scales than large ones. This scaling pattern reflects the true picture of spatial heterogeneity.

## 4.2 Detecting the scaling law

What were claimed to be power laws in the literature could be actually lognormal, exponential, or other similar distributions, because the detection of power laws can be very tricky. Given a power law relationship y=x^a, it can be transformed into the logarithm scales, i.e., $\ln(y) = a \cdot ln(x)$, indicating that the logarithms of the two varaibles x and y have a linear relationship. Conventionally, an ordinary least squres (OLS) based method was widely used for the detection. In the fractal literature, the box-counting method is usually used to compute the fractal dimension, which is the de fact power law exponent, and its computation is also based on OLS. There are at least two issues surrounding the power law detection. The first is that the OLS based method is found to be less reliable for detecting a power law, so a maximum likelihood method has been developed (Clauset et al. 2009). It was found that many claims on power laws in the literature are likely to be lognormal or other degenerated formats such as a power law with an exponential cutoff. For the sake of readability, this paper does not cover mathematical details on heavy-tailed distributions and their detection; interested readers can refer to Clauset et al. (2009) and the references therein. The second is that even with the OLS-based method, the definition of fractal dimension is so strict that many geographic features are excluded from being fractal (Jiang and Yin 2014). Given the circumstance, the authors have recently provided a rather relaxed definition of fractals, i.e., a geographic feature is fractal if and only if the scaling pattern of far more small things than large ones recurs multiple times. The number of times plus one is referred to as ht-index (Jiang and Yin 2014), an alternative index of fractal dimension, for characterizing complexity of fractals or geographic features in particular.

The idea behind the relaxed definition of fractals, or the ht-index, is pretty simple and straightforward. It is based on the head/tail breaks (Jiang 2013a), a new classification scheme for data with a heavy-tailed distribution. Given a variable whose distribution is right skewed, compute its arithmetic mean, and subsequently split its values into two unbalanced parts: those above the mean in the head, and those below the mean in the tail. The values above the mean are a minority, while the values below are a majority. The ranking and breaking process continues for the head part progressively and iteratively until the values in the head no longer meet the condition of far more small things than large ones. This way both the number of classes and the class intervals are naturally and automatically derived based on the inherent hierarchy of data. Eventually, the number of classes, or equivalently the ht-index, indicates hierarchical levels of the values. One can simply rely on an Excel sheet for the computation of the ht-index. As an example, Figure 3 illustrates the scaling pattern of the US cities, discussed earlier in Section 2.1, and it has the ht-index of 7.

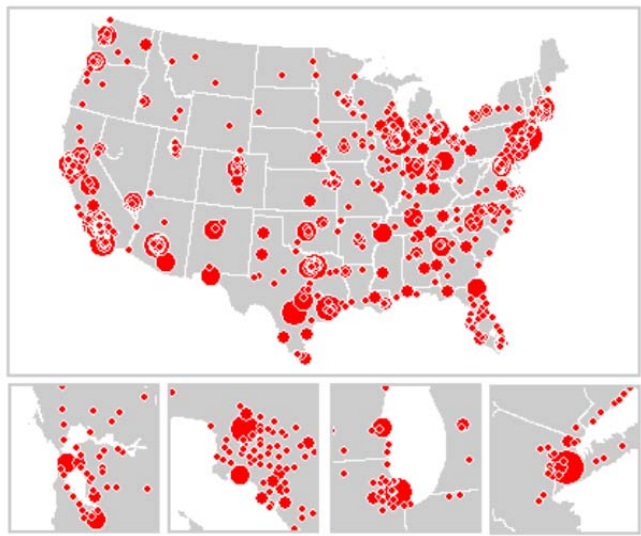

Figure 3: (Color online) Scaling pattern of US cities with ht-index equal 7
(Note: The four insets from the left to the right provide the enlarged view respectively for San Francisco, Los Angeles, Chicago and New York regions)

## 4.3 Revealing the scaling pattern

The head/tail breaks can effectively reveal or visualize the scaling pattern if the data itself exhibits a heavy-tailed distribution. This is because the head/tail breaks was developed initially for revealing the inherent scaling hierarchy or the scaling pattern. In this regard, conventional classification methods, mainly guided by Gaussian thinking, failed to reveal the scaling pattern. For example, the most widely used classification natural breaks (Jenks 1967), which is set as a default in ArcGIS, is based on the principle of minimizing within-classes variance, and maximizing between-classes variance. It sounds very natural. In some case like the US cities, the classification result of the natural breaks may look very similar to the one as shown in Figure 3. However, this is just by chance. Essentially, the natural breaks is motivated by Gaussian thinking; each class is characterized by a well-defined mean with a limited or minimized variance. In a contrast, the head/tail breaks is motivated by Paretian thinking, and for data with a Paretian-like or heavy-tailed distribution. The iteratively or recursively defined averages are used as meaningful cutoffs for differentiating hierarchical levels.

The reader probably has got used to the US terrain surface (Figure 4a), which is based on the natural breaks. It is commonly seen in geography and cartography textbooks and atlases. However, I want to challenge this conventional wisdom, arguing that the natural breaks based visualization is little natural. I contend that the head/tail breaks derived visualization is more natural, since it reflects the underlying scaling pattern of far more small things than large ones (Figure 4b). The things here are referred to individual locations, or more specifically, far more low locations than high locations. The left visualization, which distorted the scaling pattern, appears having far more high locations than the visualization to the right, or equivalently far more high locations than what it actually has. The visualization to the right reflects well the underling scaling pattern. This can be further seen from the corresponding histograms of the individual classes of the two classifications (Figure 4c and 4d). What is illustrated by the left histogram is "more low locations than high ones" which is a linear relationship, rather than "far more low locations than high ones", which is a nonlinear relationship. For the left histrogram, each pair of the adjacent bars from left to right does not constitute an unbalanced contrast of majority versus minority. For example, the first pair of bars of the left histogram shows a well-balanced contrast of 7 to 6; in a contrast, the first pair of the right histogram is unbalanced, 14 to 5. Therefore, the right histogram indicates clearly "far more low locations than high ones". Interestingly, the scaling pattern remains unchanged with respect to different scales of digital elevation models (Lin 2013).

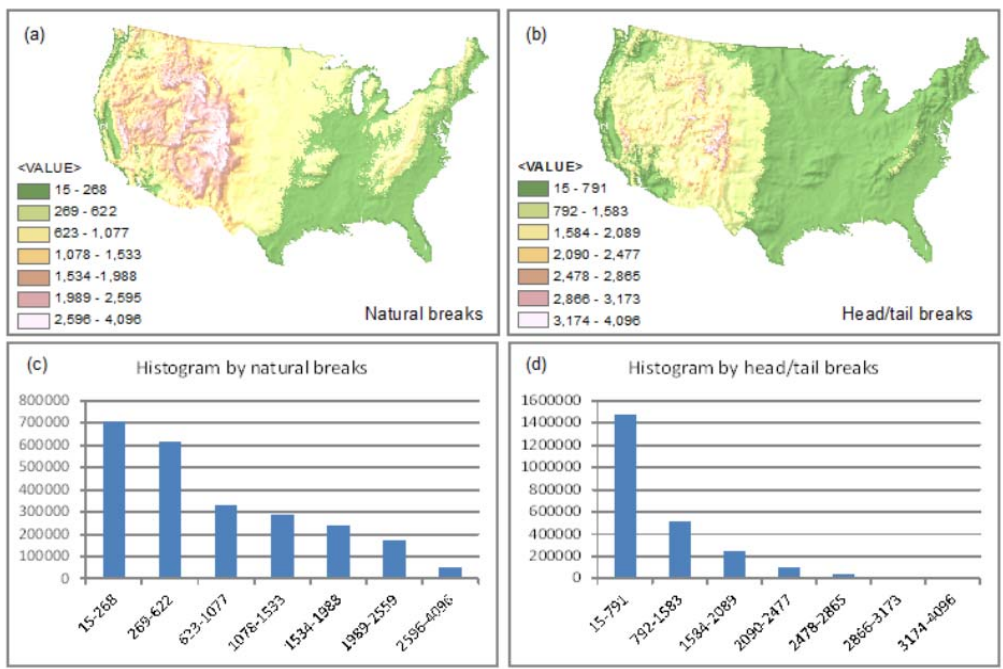

Figure 4: (Color online) The scaling pattern of US terrain surface is distorted by the natural breaks, but revealed by the head/tail breaks

## 5. Implications of Paretian thinking and the scaling law

Current geospatial analysis concentrates more on geographic forms, but less on why the forms. The forms illustrated are mostly limited to whether they are random, or to what extent they are auto-correlated. As to why the forms, it is usually ended up with simple regressions and causalities. This way of geospatial analysis is much like short-term weather forecasting. Despite its usefulness, the short-term weather forecast adds little to understanding the complex behavior of weather - the long-term weather beyond two or three weeks. In essence, the long-term weather, , or climate change in general, is unpredictable, just like earthquakes and many other events in nature and society (Bak 1996). If the real world is unpredictable, what can we do as scientists? We can simulate interactions of things from the bottom up in order to understand the underlying mechanisms, which would help improve predictions. In this regard, the emerging social media, in particular location-based social media, provide valuable data for validating the simulation results (Jiang and Miao 2014). The data, unlike traditional statistical or census data that are mainly aggregated, are not only big in size, but are collected at individual levels. The data are not only at individual levels, but linked in time and among individuals. The data can help track the trajectories of individuals and their associations in space and over time. For this kind of social media data, the scaling law and fractals should be the norm.

Current spatial statistics constrained by Gaussian thinking show critical limitations for analyzing or getting insight into big data (Mayer-Schonberger and Cukier 2013). What are illustrated by spatial statistics, either patterns or associations, can be compared to the mental images of the elephant in the minds of the blind men. These images reflect local truths, and are indeed correct partially, but they did not reflect the whole of the elephant. Geospatial analysis should go beyond illustrating spatial autocorrelation, either globally or locally, but towards uncovering the underlying scaling or fractal patterns. Geographic features are essentially and ultimately scaling or fractal. Therefore, any patterns deviating from the scaling pattern or that can be characterized by a well-defined mean are either wrong or biased.

Geographic forms (or phenomena) are not the outcomes of simple processes but the results of complex processes with positive feedbacks. In the built environments, human interventions (interdependence and interactions) of various kinds are the major effects of spatial heterogeneity. As famously stated by Winston Churchill (1874-1965), we shape our buildings, and thereafter they shape us. This statement should be comprehended in a progressive and recursive manner. This comprehension, which underlies Paretian thinking, is essentially a complex system perspective for exploring the underlying processes related to geographic forms (e.g., Benguigui and Czamanski 2004, Blumenfeld-Lieberthal and Portugali 2010). In this regard, complexity science has developed a range of tools such as discrete models, complex networks, scaling hierarchy, fractal geometry, self-organized criticality, and chaos theory (Newman 2011). All these modeling tools attempt to reveal the underlying mechanisms, linking surface complex forms (or complexity) to the underlying mechanisms (or deep simplicity) through simulations from the bottom up, rather than simple descriptions of forms or of geographic forms in particular.

Paretian thinking represents a paradigm shift. Shifting from the sands to the avalanches (Bak 1996), and from the street segments to the natural streets (Jiang 2009), enable us to see something interesting and exciting, i.e., from the things of limited sizes to the things of all sizes. The things of all sizes imply a scaling pattern across all scales. Recognition of the scaling pattern helps us to better understand the underlying universal form of geographic features. This scaling pattern can further be linked to the underlying geographic processes that are dynamic, nonlinear, and bottom-up in nature. This view would position geography in the family of science, since we geographers are interested in not only what things look like (the forms) but also why things look that way (the processes). Spatial heterogeneity is thus not a problem but an underlying scaling law of geography.

## 6. Concluding summary

This paper argues that geospatial analysis requires a different way of thinking, or world view in general, that underlies the Paretian-like distribution of geographic features. I put the two distinct views in comparison: more or less similar things in a simple, static, and equilibrium world on the one hand, and far more small things than large ones in a complex, dynamic, and non-equilibrium world on the other. Geospatial analysis has been dominated by Gaussian statistics with a well-defined mean for characterizing spatial variation ('mild' variance so to speak). Despite its ubiquity in geography, the Paretian-like heavy-tailed distribution, or the underlying way of thinking in general, has not been well received in geospatial analysis. The current geospatial analysis mainly focuses on how spatial variation deviates from a random pattern, and measuring spatial auto-correlation from global to local (the current spatial heterogeneity), but leaves the underlying processes unexplored. This way of geospatial analysis is inadequate for understanding geographic forms and processes, in particular while facing increasing amounts of social media data.

No average location exists on the Earth's surface. Instead of more or less similar things, there are far more small things than large ones in geographic space; small things are a majority while large things are a minority. Importantly, the pattern of far more small things than large ones recurs multiple times (Jiang and Yin 2014). This recurring scaling pattern reflects the true image of spatial heterogeneity that lacks a well-defined mean ('wild' variance so to speak). Spatial heterogeneity is indeed a problem in Gaussian thinking, but it is a law or scaling law in Paretian thinking. In the spirit of Paretian thinking and the scaling law, geospatial analysis should seek to simulate individuals and individual interactions from the bottom up rather than simple correlations and causalities. In this connection, complexity tools such as complex networks, agent-based modeling, and fractal/scaling provide effective means for geospatial analysis of complex geographic phenomena.

**Acknowledgement**
The author would like to thank the anonymous referees and the editor Daniel Z. Sui for their valuable comments. However, any shortcoming remains the responsibility of the author.